Friday, 9th of August 2019

# Early reduced dopaminergic tone mediated by D3 receptor and dopamine transporter in absence epileptogenesis

**Abbreviated title:** *Dopamine transmission during epileptogenesis*

**Author names and affiliations:**
Fanny Cavarec[1], Philipp Krauss[1,2], Tiffany Witkowski[1,3], Alexis Broisat[4], Catherine Ghezzi[4], Stéphanie De Gois[5], Bruno Giros[5,6], Antoine Depaulis[1*], Colin Deransart[1*]†

[1]Univ. Grenoble Alpes, Inserm, U1216, CHU Grenoble Alpes, Grenoble Institut Neurosciences, 38000 Grenoble, France

[2]Department of Neurosurgery, Klinikum rechts der Isar, Ismaninger Strasse 22, 81675 Munich, Germany

[3]Université Clermont Auvergne, INSERM U1240, Imagerie Moléculaire et Stratégies Théranostiques, F-63000 Clermont Ferrand, France

[4]INSERM-UMR INSERM U1039, Radiopharmaceutiques Biocliniques, F-38000 Grenoble, France

[5]Neuroscience Paris Seine, INSERM UMRS 1130/CNRS UMR8246, Sorbonne University, Paris, France

[6]McGill University, Graham Boeckh chair in Schizophrenia, Dpt of Psychiatry, Douglas Hospital, Montreal, QC, CAN

*These authors contributed equally to this work.

†To whom correspondence should be addressed:
Colin DERANSART, PhD
**Grenoble – Institut des Neurosciences**
INSERM U1216-UGA-CEA-CHU
Equipe 9 *Synchronization and Modulation of Neural Networks in Epilepsy*
Bâtiment Edmond J. Safra, Chemin Fortuné Ferrini
38706 La Tronche Cedex – France
Tel.: +033 (0)4 56 52 06 60
E-mail: colin.deransart@univ-grenoble-alpes.fr
†**ORCID number**: 0000-0002-9214-433X

Number of text pages: 30.
Number of words in the:
Summary: 316; Keypoints: 69; Introduction: 496; Results: 2187; Discussion: 1352.
Number of references: 50; Number of figures: 5;
Supporting information: Materials and Methods (16 pages) + Tables S1-S4 (2 pages).

**Acknowledgements:**
This work was supported by a grant from Agence Nationale pour la Recherche (BasalEpi , ANR-R06-275-CS). We thank Véronique Riban for her participation in the nafadotride experiments. We are also grateful to Hervé Mathieu, Tanguy Chabrol, Carole Carcenac, Hélène Bernard, Anne Bertrand, Isabelle Guillemain, Céline Beaumont and Pierrick Bordiga for their technical assistance. We greatly acknowledge the valuable advices from Jakub Toczek, Marcelle Moulin and Philippe Millet for the DATScan analysis, S. Ananthan for the generous gift of SR21502 and P. Sokoloff for the generous gift of nafadotride. MRI facility IRMaGe was partly funded by the French program "Investissement d'Avenir" run by the 'Agence Nationale pour la Recherche'; grant 'Infrastructure d'avenir en Biologie Santé' – ANR-11-INBS-0006. F.C. received a PhD fellowship from the Ministère Français de la

**Abbreviations**

| | |
|---|---|
| 5HIAA | 5-hydroxyindolacetic acid |
| 5HT | serotonin |
| APZ | Aripiprazole |
| D3R | Dopamine D3 receptor |
| DA | Dopamine |
| DAT | Dopamine transporter |
| DOPAC | 3,4-dihydroxyphenylacetic acid |
| GAERS | Genetic Absence Epilepsy Rats from Strasbourg |
| HVA | Homovanillic acid |
| Acb | Accumbens nucleus |
| AcbC | Accumbens nucleus, core |
| AcbSh | Accumbens nucleus, shell |
| NEC | Non epileptic controls |
| SWD | Spike-and-wave discharges |
| TO | Olfactory tubercles |

**Summary**

**Objective**: In Genetic Absence Epilepsy Rats from Strasbourg (GAERS), epileptogenesis takes place during brain maturation and correlates with increased mRNA expression of D3 dopamine receptors (D3R). Whether these alterations are the consequence of seizure repetition or contribute to the development of epilepsy remains to be clarified. Here, we addressed the involvement of dopaminergic system in epilepsy onset in GAERS.

**Methods**: Experiments were performed using rats at different stages of brain maturation from three strains according to their increasing propensity to develop absence seizures: Non Epileptic Controls rats (NEC), Wistar Hannover rats (WH) and GAERS. Changes in dopaminergic neurotransmission were investigated using different behavioral and neurochemical approaches: autoradiography of D3R and dopamine transporter, SPECT imaging, acute and chronic drug effects on seizure recordings (dopaminergic agonists and antagonists), quinpirole-induced yawns as well as dopamine synaptosomal uptake, microdialysis, brain tissue monoamines and BDNF quantification.

**Results**: Autoradiography revealed an increased expression of D3R in 14-days old GAERS, before absence seizure onset, that persists in adulthood, as compared to age-matched NEC. This was confirmed by increased yawns, a marker of D3R activity, and increased seizures when animals were injected with quinpirole at low doses to activate D3R. We also observed a concomitant increase in the expression and activity of the dopamine transporter in GAERS before seizure onset, consistent with both lowered dopamine basal level and increased phasic responses.

**Significance**: Our data show that the dopaminergic system is persistently altered in GAERS, which may contribute not only to behavioral comorbidities but also as an etiopathogenic factor in the development of epilepsy. They suggest that an imbalanced dopaminergic tone may participate in absence-epilepsy development and seizure onset as its reversion by a chronic treatment with a dopamine stabilizer significantly suppressed epileptogenesis. Our data propose a potential new target for antiepileptic therapies and/or improvement of quality of life of epileptic patients.

**Key points (3-5 bullets, less than 140 characters each)**

- Dopamine neurotransmission is well known to modulate epileptic seizures in animal models.
- Increased dopamine transporter and D3R receptor expression and activity already occur before the appearance of seizure in absence epilepsy prone rats.
- These changes likely result in both lowered basal level of dopamine and increased phasic responses.
- Imbalance in dopaminergic tone may participate in epilepsy onset and seizure occurrence during epileptogenesis, as well as in to behavioral comorbidities.

## Introduction

Dopamine (DA) has long been associated with epilepsy.[1] Indeed, drugs that activate DA neurotransmission have antiepileptic effects, whereas those – like antipsychotics – that decrease DA neurotransmission are well known to aggravate seizures.[2] Following the first PET-scan evidence of a dopaminergic dysfunction in ring chromosome 20 epileptic patients,[3] several clinical studies have reported imaging evidence of DA neurotransmission alterations in a variety of epileptic syndromes.[4,5] Altogether, these data showed that a reduction of dopaminergic activity is associated with recurrent seizures and suggested that this neurotransmission may play a role in the physiopathology of idiopathic epilepsies. However, these clinical observations, in line with reports in animal models of chronic epilepsy, were observed while the epileptic syndromes were fully developed and therefore could rather be considered as a consequence of seizure repetition.[6,7] Using a well-recognized model of absence epilepsy in the rat, the Genetic Absence Epilepsy Rat from Strasbourg (GAERS),[8] we addressed the possibility that a constitutive dopaminergic dysfunction may participate in the building up of seizures (i.e., epileptogenesis) by favoring their progressive occurrence and aggravation. In this model, low-amplitude epileptiform oscillations first occur in the somatosensory cortex at 14 days (P14) and progressively evolve as spike-and-wave discharges (SWD) – i.e., the electrophysiological hallmark of absence seizures – which are recorded not before post-natal day 25 (P25).[9] The number of spike-and-wave patterns per discharge as well as the duration of discharges then progressively increase, to stabilize in adults after post-natal day 90 (P90).[9]

The use of GAERS to address a possible involvement of dopaminergic neurotransmission during epileptogenesis is further justified by the fact that, as in human patients, SWD are modulated by dopaminergic ligands. In GAERS, systemic injections of D1 and D2 receptor agonists suppress absence-seizures, whereas receptor antagonists aggravate them.[10] Similar effects are observed when these compounds are directly injected in the ventral striatum – namely the core of the nucleus accumbens (AcbC).[11] Among the different transcripts involved in dopaminergic neurotransmission, we reported that only D3 receptor mRNAs are over-expressed in adult GAERS AcbC, as compared to age-matched Non-Epileptic Control rats (NEC, an inbred strain of rats without seizures derived from the same original Wistar Hannover (WH) rat colony).[6] In addition to its postsynaptic role, D3 receptor (D3R) also acts as an autoreceptor and inhibits DA release and synthesis leading to a control of phasic activity of dopaminergic neurons.[12] Furthermore, D3R expression varies during the first weeks of age in rats, suggesting a functional role in cerebral maturation.[13,14] Finally, *in vitro* experiments have suggested that D3R regulate the dopamine transporter (DAT) and it is therefore possible that the increase of D3R expression observed in GAERS could impact on the expression and/or functionality of DAT.[15,16] In the present study, we examined whether an increase of D3R expression in juvenile GAERS could participate in absence epileptogenesis and lead to a sustained reduced dopaminergic tone in these animals directly and/or through a regulation of DAT expression and/or function. We then examined the consequences of this dysregulation in the control of SWD in adult animals.

Friday, 9th of August 2019

## Materials and methods

All materials and methods are described in online Appendix S1

## Results

See Tables S1 to S4 in online Appendix Tables

### *1. Increased D3R expression in GAERS before seizure onset*

We first explored whether D3R expression was increased during epileptogenesis in GAERS. Using [$^{125}$I]-PIPAT autoradiography, we compared D3R densities in selected regions of interest at 14, 21 and 90 postnatal days in GAERS to two age-matched control strains (WH and NEC). At P14, we observed a significant increase of D3R expression in striatum and AcbC in GAERS, as compared to control rats (Fig. 1, A1-2). At this age, we also observed a clear D3R expression in the granular layer of the barrel field of somatosensory cortex and in the ventrobasal thalamus (results not shown) in the three strains, in agreement with Gurevich and Joyce, 2000.[14] D3R expression in the barrel field of somatosensory cortex was significantly higher in GAERS as compared to NEC (Table S1). At P21, D3R expression remained higher in GAERS when compared to NEC in most structures analysed (Table S1). At P90, D3R expression was mainly increased in AcbC, striatum and olfactory tubercles, as compared to NEC (Fig. 1, A1-2; Table S1). Beside significant differences in the expression levels between GAERS and NEC, whatever their age, the differences in expression levels between GAERS *vs* WH, and NEC *vs* WH were relatively weak and generally did not reach significance. Nevertheless, there was a general trend for WH to systematically display intermediate levels between GAERS and NEC.

Because BDNF was shown to control D3R expression in adults,[17] we hypothesized that the higher level of D3R expression observed in GAERS could be triggered by an increased BDNF expression. However, using Elisa quantification, we did not find any significant differences in BDNF protein levels among the three strains at P90 (Table S2).

Altogether, these data showed, whatever the age, a higher expression of D3R in GAERS, in particular in the striatum and the AcbC. This higher expression was not associated with an increase of BDNF.

## 2. Increased D3R activity in GAERS before seizure onset

Because splice variant isoforms of D3R were suggested to produce non-dopamine binding proteins,[18] it was necessary to confirm that the increase of D3R expression in GAERS had functional consequences. To verify this point, we used yawning as a D3 activation behavioral assay. Indeed, yawns have been shown to be specifically induced by systemic injection of low doses of quinpirole (a D2/D3R agonist) and to depend upon D3R activation in both animal models and humans.[19] This paradigm was further justified by our previous observations of a decrease in spontaneous yawns in adult GAERS as compared to NEC.[20]

The basal rate of yawns/hour, whatever the age and strain, was not different before and after s.c. injections of saline (data not shown). After s.c. saline injections, we observed that GAERS at P14, P21 and P90, displayed significantly less spontaneous yawns than control rats (Fig. 1 B1). On the contrary, after quinpirole injection (12.5, 25 and 50 µg/kg, s.c., for P14, P21 and P90, respectively – doses chosen to avoid D2R-mediated motor side effects), GAERS displayed significantly more yawns than NEC or WH at any ages (Fig. 1 B2). A similar increase obtained in adults with the D3 receptor agonist PD128907 (results not shown). In order to control for a possible confounding motor side-effects following quinpirole injections, the number of induced-rearings was also quantified after each quinpirole injection. However, this quantification did not revealed any differences (data not shown).

Altogether, these data showed, whatever the age, a lower basal rate of yawns/hour and a higher expression of quinpirole-induced yawns in GAERS as compared to the NEC and WH. They confirmed a higher activity of D3R in GAERS already before seizure onset.

### 3. No changes in DAT expression in adult GAERS

*In vitro* experiments have suggested that D3R regulate DAT function and it is therefore possible that the higher expression of D3R observed in GAERS could lead to a change in DAT expression and/or function.[15, 16] We first explored this possibility in P90 GAERS as compared to NEC and WH by *in vivo* SPECT imaging with [$^{123}$I]-Ioflupane (DaTSCAN$^{TM}$). We found no significant differences between strains in the striatum, AcbC or olfactory tubercles (Fig. 2, A2). Immediately after *in vivo* imaging, we performed *ex vivo* autoradiography using the remaining presence of radioactive tracer in the brain. Although a trend to increased autoradiogram signals was observed in GAERS, we found no significant difference in DAT expression between the three strains (Fig. 2, A3).

Because DaTSCAN$^{TM}$ could be reliably performed only in adult P90 rats, we used [$^{3}$H]-GBR12935 autoradiography on slices in P14, P21 and P90 animals to further explore DAT expression.[21] At P14, we observed that the mean optical density of [3H]-GBR12935 binding was significantly higher in striatum and AcbC (Fig. 2, B1-B2) and in olfactory tubercles of GAERS (Table S3). This increase was significant only versus NEC at P21 in the AcbC. At P90, we observed no differences between strains in any structure (Fig. 2, B1-2), in agreement with our *in vivo* SPECT imaging data.

Altogether these data showed that DAT expression is only slightly increased in P14 GAERS as compared to NEC and WH, and that this difference vanishes thereafter in adults.

### *4. Increased DAT activity in GAERS before seizure onset*

To precisely determine whether the above changes in DAT expression level were indicative of modifications in functional activity, we directly measured [3H]-DA uptake in synaptosomal striatal fractions in P14 and P90.

In P14 rats, we observed a significant increase of DAT-mediated [3H]-dopamine uptake in GAERS compared to WH or NEC (Fig. 3A). This corresponds to an increase in dopamine translocation velocity with an estimated $V_{max}$ for DAT-mediated $^3$H-DA uptake which was increased by 25-30% (GAERS: 8.26 ± 0.47 pmol/mg/min; NEC: 6.25 ± 0.44 pmol/mg/min; WH: 6.58 ± 0.16 pmol/mg/min; *p*<0.025 and p<0.05 respectively, t-test; Fig. 3C). No significant changes in DA affinity for the transporter were observed between strains (GAERS $K_M$: 268.9 ± 57.2 nM; WH $K_M$: 290.6 ± 25.3 nM; NEC $K_M$: 261.1 ± 68.7 nM; Fig. 4C). In P90 rats, DAT activity was increased in GAERS as compared to NEC (+17%; Fig 3B), due to an increase in $V_{max}$ (GAERS: 16.61 ± 0.55 pmol/mg/min; NEC: 14.17 ± 0.64 pmol/mg/min *p*<0.01, t-test; Fig. 3D) with no change in DA affinity (Fig. 3F).

However, we observed no significative differences between GAERS and WH at this age (GAERS: 16.61 ± 0.55 pmol/mg/min; WH: 17.2 ± 0.63 pmol/mg/min). These data suggest that enhanced DAT activity is unlikely due to changes in intrinsic properties of DAT (turn-over rate or recognition of DA for DAT ligand binding domain) but rather to an increase in the number of transporters expressed at striatal plasma membrane. This is in line with an increased dopamine release induced by 2-µM amphetamine that we observed in adult GAERS synaptosomes as compared to NEC ones (Fig. 3C).

Altogether, these results show that changes in DAT function (increased translocation velocity) also occur early during epileptogenesis in GAERS.

## 5. *No changes in tissular DA levels in adult GAERS*

Our data reported above suggested that the increase in D3R expression and function and DAT activity during epileptogenesis should both lead to a decreased dopaminergic tone in adult GAERS, as compared to NEC and WH. To examine this hypothesis, we first measured tissular levels of monoamines and their metabolites by HPLC from punches of tissues performed in P90 animals in different structures. Beside a slight but significant decrease in DA levels in the striatum in NEC *vs* WH (Table S4), we found no significant changes between the strains. Similarly, no changes were observed in DOPAC, HVA, 5HT or 5HIAA concentrations. Furthermore, we found no differences in the DA turn-over as determined by DOPAC/DA or HVA/DA ratios, suggesting no change in DA metabolism between the three strains. Similarly, no differences were observed in 5HIAA/5HT concentration ratios (results not shown).

Altogether, we found no changes in *post-mortem* levels of monoamines between the three strains suggesting that the changes reported above during brain maturation/epileptogenesis did not lead to gross alteration of DA metabolism. However, such tissular measurement of monoamines levels could not provide any dynamic information on the extracellular DA that is directly involved in neurotransmission.

## 6. Lower basal extracellular DA levels and increased dopaminergic reactivity in GAERS as revealed by in vivo microdialysis

In this experiment and the followings we decided to focus on the comparisons between GAERS and NEC to avoid the possibility that occurrence of occasional spontaneous SWD in WH interfere with our measurements.[22]

Using microdialysis, we observed that the basal extracellular content of dopamine collected in AcbC for 90 min was 52% lower in GAERS than in NEC (Fig. 4A).

Because amphetamine induced a greater DA release in GAERS synaptosomes (see above), we hypothesized that its systemic injection would also result in a greater increase in DA release *in vivo*. Prior to microdialysis experiments, we tested the effects of different doses of amphetamine (0.3, 0.6 and 0.9 mg/kg, i.p.) on the occurrence of SWD in a different group of adult GAERS and observed a dose-dependent suppression of cumulated duration, number and mean duration of SWD for 80 min (Fig. 4B), in agreement with a previous report.[10] At the highest dose, however, increased exploration of the test cage, sniffing, twitching of the vibrissae, as well as rearing were observed whereas no such behavioral effects were seen at lower doses (data not shown). We therefore decided to use the 0.6 mg/kg dose in microdialysis experiment. When challenged with such a dose, GAERS displayed a higher increase in extracellular DA levels as compared to NEC, which was significant for 60 min post-injection (Fig. 4C). The time-course of this effect was in agreement with SWD suppression observed above and no changes in animals' general behavior was observed. Similarly, a 15-min high concentration KCl perfusion through the microdialysis probe performed 150 min after the injection of amphetamine, resulted in a higher increase in DA levels in GAERS than in NEC, which was significant for 20 min post-injection (Fig. 4C). No change was observed in the levels of other monoamines and metabolites (data not shown).

Altogether, these results showed a lower basal extracellular DA level in the AcbC of adult GAERS as compared to NEC, but an increased release when stimulated by systemic amphetamine or locally applied KCl.

## *7. Effect of D3R modulation on the occurrence of absence seizures in adult GAERS*

Because D3R activation reduces DA release[23] and since our previous and present data showed suppression of SWD when DA release is increased, we hypothesized that D3R activation should increase the occurrence of SWD in adult GAERS. To test this hypothesis we used quinpirole, a DA receptor agonist with a greater affinity for D3 than for D2 receptors.[24] Using the same dose as above to elicit yawns in P90 GAERS (50 µg/kg, s.c.), we observed a significant increase of cumulated duration and number of SWD for up to 40 min (Fig. 5A1). Analysis of video-EEG recordings also revealed that about 60% of yawns occurred during SWD (of which 83% interrupted the on-going seizures) whereas 40% occurred during inter-ictal periods. When the same dose was injected to NEC or WH rats in a previous study, no SWD or epileptiform events were observed in NEC, whereas 3 out of 8 WH rats displayed sparse seizure-like activities.[20] This proepileptic effect was confirmed with (+)-PD-128,907, a compound more specific to D3R[23] (Fig. 5A2). Of note, (+)-PD-128,907 also induced yawns (results not shown). The aggravation of SWD by the two D3R agonists, suggested that D3R antagonists should decrease their occurrence. However, we observed no significant changes with neither SR21502 nor SB277011-A, two D3R receptor antagonists at doses known to inhibit D3R (Fig. 5, B1-2).[25,26] Similarly, the D3R antagonist nafadotride[27] had also no effect on SWD occurrence and duration (data not shown).

Altogether these data showed that D3R agonists displayed proepileptic effects in GAERS, whereas D3R antagonists had no effect on SWD occurrence.

## 8- Suppression of absence epileptogenesis by chronic treatment with a mood stabilizer

In a last experiment, we addressed the hypothesis that an imbalance of the dopaminergic tone (i.e., lower DA basal level and higher reactivity to amphetamine or KCl) would be critical in SWD occurrence and/or epileptogenesis. To this aim, we used chronic administration of aripiprazole (APZ), a dopamine stabilizer. Although APZ has a complex pharmacological profile, this compound has "functionaly selective" actions on D3 and D2R coupled with important interactions with selected other biogenic amine receptors, particularly at 5HT2B, 5HT1A and 5HT2A receptors.[28] APZ was shown to increase tonic DA release and to decrease phasic DA release upon chronic administration.[29] As predicted by our previous experiments with D3R agonists, acute injection of APZ (3, 5 and 10 mg/kg, i.p.) in P90 GAERS significantly increased the number and cumulated duration of SWD in a dose-dependent manner, with significant effect at 5 and 10 mg/kg (Fig. 5, C1). By contrast, chronic administration of APZ (3 mg/kg/day, s.c.) in GAERS rats (n=9) from birth (P0) to P90 significantly reduced the number of SWD by an average of 30% at all time-points (Fig. 5, C2). During this treatment, we observed no obvious differences in rats weight, general behavior or motor coordination (assessed by rotarod), as compared to controls (data not shown). Since APZ has also been described as a DA D3R partial agonist, its chronic administration is likely to modify D3R transcripts expression[30]. Therefore, at the end of the APZ treatment (P90), rats were killed and we perform [$^{125}$I]-PIPAT autoradiography. We found no significant difference in D3R expression in the striatum (Fig. 5, C4) but a significant decrease in AcbC (Fig. 5, C5), in APZ-treated compared to vehicle-treated GAERS. These data support our hypothesis that an imbalance in dopaminergic tone plays a critical role in absence epileptogenesis in GAERS and may be, at least in part, mediated by D3R.

# Discussion

An abnormal balance in tonic and phasic DA release has been suggested to play a critical role in the pathophysiology of several neurological and psychiatric disorders.[12,31] Our data are in line with this concept and suggest that an early imbalance of DA release during brain maturation in conjunction with an abnormal network maturation process, may participate in absence epileptogenesis and, later on, in the maintenance of seizure susceptibility. Our results suggest that it could be related to changes in D3R and DAT expression and/or activity.

**DA tone during the maturation of the epileptic network**

Overexpression of D3R already appears in GAERS pups, before SWD onset, which is confirmed by an increase in quinpirole-induced yawns. However, whatever their age, GAERS display almost no spontaneous yawns as compared to NEC and WH. These results suggest that the tonic release of DA in GAERS is not sufficient to allow for a basal expression of yawns, in line with their decreased levels of DA measured by microdialysis. Similarly, although the overall DAT density was not modified in adult GAERS, its transient overexpression in pups and weanlings may persist at the neuronal membrane, as evidenced by DA uptake experiments in synaptosomes. Such changes may lead to an increased clearance of DA and contribute to a decreased dopaminergic tone before the onset of absence-epilepsy. Our data are in disagreement with the 40-50% decrease in DAT expression that was reported in the caudate nucleus and AcbC of adult GAERS[32], a difference that might be rather due to several drug pre-treatments according to the experimental design[32] and/or to the genetic drift in the GAERS colony used for this study.[33]

It is noteworthy that immature SWD first occur once the pruning of DA cell population has been completed (i.e., by P20) to reach the final adult population of midbrain DA neurons.[9,31] In GAERS, the early increase in D3R and DAT expressions as soon as P14 suggests that changes in the DA tone could be associated with an abnormal pruning of the DA system. Importantly, the overexpressed D3R are mainly seen within the deepest layers of the somato-sensory cortex and may therefore influence the maturational process of this cortical region shown to initiate SWD.[9] Indeed, a different development of the dendritic arbor of the barrel neurons may participate in the propensity of the somatosensory cortex to develop absence epilepsy.[14] Although no difference was found in BDNF expression between strains in adults, the possibility remains that a transient difference would occur during brain maturation.[24]

The persistent changes in dopaminergic tone that we observed in GAERS pups would favor the emergence of abnormal epileptiform thalamo-cortical oscillatory activities and the progressive build-up of SWD, in addition to the abnormal maturation of the cortical network suggested by our previous study.[9] These changes may impair the functionning of the basal ganglia network known to exert an inhibitory remote control on epileptiform cortical activities.[34] Hence, the recurrence of SWD may aggravate the imbalance in dopaminergic tone by maintaining an increased D3R expression and DAT activity, therefore reducing over-fluctuations of DA levels. This hypothesis is further supported by the progressive increase of SWD occurrence and duration during the first 3-4 months of life in GAERS.[9] A decreased dopaminergic tone during this period may thus participate to the progressive aggravation of seizures. This hypothesis is in agreement with the reduction of SWD observed in GAERS (present study) and in WAG/Rij rats after a chronic treatment with APZ.[35] Indeed, this reduction is associated with a decreased D3R expression in GAERS, in line with a previous report using chronic treatment with a neuroleptic.[36] Furthermore, chronic APZ was reported to improve epileptic disorders in humans,[37] therefore supporting that modifying dopaminergic tone may slow-down the development of epilepsy.[38] Given the complex pharmacological profile of APZ, it cannot be ruled out, however, that the effects on other biogenic amines (e.g., 5-HT) also contribute to the chronic reduction of seizures.[28]

**Decreased dopaminergic tone and ictogenesis**

D3R activation by systemic injection of an agonist increased SWD in adult GAERS in the present study but never induced any seizure when injected in NEC.[20] D3R agonists were reported to increase epileptiform activities in Wistar rats in which occasional spontaneous SWD can be observed, and high-voltage low-frequency EEG waves in rabbits.[20,39] This aggravating effect appears specific to D3R receptors since activations of D1 or D2 receptors suppress SWD in GAERS.[10,11] Conversely, we found that D3R blockade had no effects on seizures. This is in agreement with the fact that most D3R antagonists have no effects *per se* on DA efflux.[40,41] This is also in line with the concept that D3R activation exerts a phasic, but not tonic, control on DA neurons activity.[12] Of all DA receptors, D3R have the highest affinity for DA and therefore plays a prominant role in the control of brain DA levels.[24] For instance, an increased basal level of DA was observed in $D3R^{(-/-)}$ mice[42] where a delayed onset of clonic seizures and increased sensitivity to the anticonvulsant effects of diazepam were reported.[43] In line with these data, the higher expression of D3R in GAERS should hamper seizure-related phasic DA over-fluctuations, contribute to lower DA tone and epileptic threshold, and therefore account for ictogenesis susceptibility.

In adult GAERS, we previously recorded a transient decrease of spiking activity in the dopaminergic neurons of the the *substantia nigra compacta* during SWD followed by an increase upon termination of the seizures.[44] This time-course suggests that a phasic increase in DA release occurs at the termination of each SWD in GAERS and may therefore influence DA equilibrium. The frequent occurrence of seizures in adult GAERS (up to 1/min when the animals are at rest) appears likely to promote an increase in DA clearance. In accordance with this hypothesis, we showed an increased DA release in GAERS when challenged by amphetamine, as well as an increase in DAT activity. These data suggest that the recurrence of SWD may aggravate the imbalance in dopaminergic tone by maintaining an increased D3R expression and DAT activity, therefore reducing over-fluctuations of DA levels, in particular during brain maturation.

**Dopaminergic influence on comorbidities and absences**

Our data, in line with clinical reports, suggest that a dysfunction of dopaminergic transmission, as well as a reduced tone mediated by D3R and DAT, are likely to be present in most epileptic syndroms.[3-5] Furthermore, several comorbidities known to be associated with a decrease of dopaminergic activity could be expressed *before* epilepsy onset.[45] Whether these changes appear before or during epileptogenesis is difficult to assess in human patients but could be addressed in animal models. For instance, adult GAERS display more defecations, less spontaneous exploratory behavior and are in general quite placid,[20] a hypodopaminergic phenotype that may account for their reduced locomotion and rearings, as compared to NEC and WH in the present study.[20,46] In addition, the decrease of REM-sleep in GAERS[20,47] could also reflect their DA-depleted state, as already demonstrated in rodents.[48] A link between the ontogeny of yawns and sleep-states has been shown, which is consistent with both decreased spontaneous yawns and REM-sleep in GAERS.[49] In WAG-Rij, a reduced dopaminergic tone was suggested to be associated with a depressive-like phenotype,[7] whereas it could account for the psychotic-like and depressive disorders reported in GAERS.[4,32] Finally, apathetic-like behaviors can be induced in rats by adenovirus-driven overexpression of D3R.[50] In reverse, D3R stimulation decreases apathy in parkinsonian patients.[51] Altogether, our data thus support that early changes in dopaminergic neurotransmission would account for the development of comorbidities in addition to their epileptic phenotype. Hence, any condition increasing phasic dopamine in the epileptic brain, including that related to activation of the mesocorticolimbic DA system that may also induces an emotional positive state (a state referring to its own prolongation and/or intensification) will increase seizures susceptibility.[52] Whether DA-associated comorbidities could be considered as biomakers of epileptogenesis remains to be further examined in animal models.

Because SWD are initiated in the somatosensory cortex, it is important to note that DA was shown to reduce primary sensorimotor cortex excitability in adult rats, thereby improving sensory signal-to-noise ratio.[53] Conversely, DA antagonists increase the excitability of somatosensory cortex to afferent signals.[53] Dopaminergic modulation may thus render somatosensory cortex circuitry more effective in processing sensory information from different sources. Our data suggest that ictogenesis is likely to be associated with a cyclic and transitory disruption in DA modulation in an abnormally oscillating network. In this context, seizures would occur during a decreased tonic release of DA which may act as a facilitating and self-sustaining factor, which may also lead to a transient alteration of consciousness via an increased excitability of somatosensory cortex to afferent signals. The seizure may thus last until an efficient attractor allows for a rebound in DA activity leading to increased phasic release of DA that will allow to be back to an efficient processing of sensory information.

**Legends**

FIGURE 1:

**D3R expression and function in GAERS, NEC and WH during epileptogenesis**. A-B: D3R density as measured by [$^{125}$I]7OH-PIPAT autoradiography in GAERS, NEC and WH, at P14, P21 and P90-days old. Data are expressed as mean optical density ± SEM in the striatum (A1) and the nucleus accumbens core (A2) (n=7 for each strain at the different ages; *$p$<0.05 *vs* NEC, #$p$<0.05 *vs* WH; Kruskal-Wallis and Dunn's test for each brain region at each ages). Induction of yawns by saline (B1) and the dopamine D3-preferring agonist quinpirole (B2; 12.5 µg/kg at P14; 25 µg/kg at P21; 50 µg/kg at P90). Data are presented as mean (± SEM) numbers of yawns during a 60-min observation period. Respectively in GAERS, NEC and WH, at P14 (n= 7, n=5, n=7), P21 (n=10, n=7, n=6) and P90 (n=6, n=5, n=7); *$p$<0.05 vs NEC, #$p$<0.05 vs WH; Kruskal-Wallis and Dunn's test at each ages.

FIGURE **2**:

**Dopamine transporter SPECT imaging in adult GAERS, NEC and WH and DAT autoradiography during epileptogenesis.** *In vivo* (A2, SPECT quantitation) and *ex vivo* (A3, autoradiographic quantitation) binding potential were investigated by DaTScan imaging. Regions of interests (ROI) (A1, left; AcbC: nucleus accumbens core; TO: olfactory tubercles) were drawn in order to quantify DaTScan uptake as determined using either *in vivo* SPECT/MRI merged images (A1, center, in a WH) or *post-mortem* autoradiographic images (A1, right, in a WH). Results were expressed as ROI to Cerebellum Ratios for three representative structures in GAERS, NEC and WH (n=6, n=5 and n=6 respectively; $p$<0.05 vs NEC and WH, Kruskal-Wallis and Dunn's test). B1-2: Dopamine transporter density as measured by [$^3$H]GBR12935 autoradiography in brain slices of GAERS, NEC and WH at P14 (n=7, n=5 and n=7), P21 (n=10. n=7 and n=6) and at P90 (n=6, n=5 and n=7), respectively. Data are expressed as mean optical density ± SEM; (*$p<0.05$ vs NEC, #$p<0.05$ vs WH, Kruskal-Wallis and Dunn's test).

FIGURE 3:

**DA uptake in synaptosomal striatal fractions** in P14 and P90 rats of the three strains. A-B: Kinetic analysis of specific ($^3$H)-DA saturation uptake rates in striatal synaptosomes at P14 and P90. 10 animals were used for each concentration and group. Values represent the means ± SEM. The kinetic parameters estimated by non-linear regression fiting of the data to the Michaelis-Menten equation.*$p<0.05$ vs NEC, #$p<0.05$ vs WH. C: Effects of 15-min préincubation with 2µM amphetamine on (3H)-DA release in P90 animals. 6 animals were used par strain. Histograms show the mean ± SEM of relative counts/minute (*$p<0.05$ vs NEC, #$p<0.05$ vs WH, Kruskal-Wallis and Dunn's test).

FIGURE 4:

**Changes in DA release as investigated by intracerebral microdialysis.** (A) Basal dopamine levels (nM) as determined from the first six samples in C (*$p<0.05$, Mann-Whitney test). (B) Effects on absence seizure of amphetamine (0, 0.3, 0.6 and 0.9 mg/kg, i.p., n=16). Data expresssed as mean (±S.E.M.) of SWD cumulated duration (*$p<0.05$ vs vehicle, Wilcoxon test). (C) Histogram of data expressed in each strain as the mean percentage (± SEM) of the corresponding dialysate outpout obtained from the first six samples (GAERS: n= 8; NEC: n=7). The whole experiment consisted in collection of 26 consecutive samples, each corresponding to a 15 min dialysate (see Materials and Methods section). Amphetamine (0.6 mg/kg, i.p.) was injected after the 6th sample and KCl infused for 15 min after the 16th sample (*$p<0.05$ *vs* NEC, 2-way RM Anova followed by Sidak's multiple comparison test). Inset: microdialysis probe implantation verification; arrowheads indicate the segment of the microdialysis membrane within the nucleus accumbens core.

FIGURE 5:

**Acute and chronic pharmacological modulation of absence-seizures.**

A-B: **Effects on absence-seizures of subcutaneous injections of compounds acting at the D3R.** Effects on absence-seizure of D3R agonists: quinpirole (A1; 50 mg/kg. s.c.. n=8); (+)-PD128907 (A2; 50 mg/kg. s.c.. n=8). Effect on absence seizure of D3R antagonists: SR21502 (B1; 7,5;15 mg/kg. s.c.. n=8) and SB277011-A (3; 9 mg/kg. s.c.. n=8). Data are presented as mean (± SEM) of SWD cumulative duration. (*$p<0.05$ versus vehicle, Wilcoxon test). C: **Effects of acute and chronic aripiprazole on absence seizure in GAERS.** Acute treatment (C1; n=8). Chronic (90-days) treatment with 3 mg/kg APZ i.p. (C2; APZ, n=9; vehicle, n=7) Significant differences from vehicle group of the same age are shown (*$p<0.05$. vs vehicle, 2-way RM Anova followed by Sidak's multiple comparison test). C4-5: Bar graphs depicting ($^{125}$I)7-OH-PIPAT binding in the striatum (C4) and in the nucleus accumbens core (C5) to the D3R of rats chronically injected with APZ until P90. Data are expressed as mean optical density ± SEM (n=7 for each strain at the different ages; *$p<0.05$ vs NEC, #$p<0.05$ vs WH, Kruskal-Wallis and Dunn's test for each brain region at each ages).

Friday, 9th of August 2019

**Statements:**

Authors confirm that they have read the Journal's position on issues involved in ethical publication and affirm that this report is consistent with those guidelines. Authors declare no conflict of interest as part of the submitted manuscript.